\documentclass[nofootinbib,aps,preprint,superscriptaddress]{revtex4}
\usepackage{amsmath} 
\usepackage{citesort}
\usepackage{epsfig}

\newcommand{\beqa}{\begin{eqnarray}}
\newcommand{\eeqa}{\end{eqnarray}}
\newcommand{\beq}{\begin{equation}}
\newcommand{\eeq}{\end{equation}}
\newcommand{\bal}{\begin{align}}
\newcommand{\eal}{\end{align}}
\renewcommand{\Re}{{\cal R}e}
\renewcommand{\Im}{{\cal I}m}

\def\gsim{\ \rlap{\raise 3pt \hbox{$>$}}{\lower 3pt \hbox{$\sim$}}\ }
\def\lsim{\ \rlap{\raise 3pt \hbox{$<$}}{\lower 3pt \hbox{$\sim$}}\ }

\def\Arhok{a_{\rho}} 
\def\Akpi{a_{K^*}}
\def\Dbar{\overline D}
\def\DbarZ{\overline{D}^0}

\def\bwotw{{\cal A}_{K^{*}}}
\def\bwoth{{\cal A}_{K^{*}}}
\def\bwtt{{\cal A}_{\rho^0}}

\begin{document}  
  
\preprint{\vbox{\hbox{hep-ph/0303187}
  \hbox{March, 2003}}}  
  
\vspace*{18pt}   
  
\title{\boldmath Determining $\gamma$ using $B^\pm \to D K^\pm$ 
with multibody $D$ decays}

\author{Anjan Giri}    
\author{Yuval Grossman}     
\affiliation{Department of Physics,     
Technion--Israel Institute of Technology,\\ 
Technion City, 32000 Haifa, Israel}
\author{Abner Soffer}
\affiliation{Department of Physics, Colorado State University, 
Fort Collins, CO 80523}
\author{Jure Zupan}
\affiliation{Department of Physics,  
Technion--Israel Institute of Technology,\\  
Technion City, 32000 Haifa, Israel}
\affiliation{J.~Stefan Institute, Jamova 39, P.O. Box 3000,1001
Ljubljana, Slovenia\vspace*{18pt}}

\begin{abstract} \vspace*{18pt}  
We propose a method for determining $\gamma$ using $B^\pm\to D K^\pm$
decays followed by a multibody $D$ decay, such as $D \to
K_S\,\pi^-\pi^+$, $D \to K_S\,K^-K^+$ and $D \to K_S\,
\pi^-\pi^+\pi^0$.  The main advantages of the method is that it uses
only Cabibbo allowed $D$ decays, and that large strong phases are
expected due to the presence of resonances. Since no
knowledge about the resonance structure is needed, $\gamma$ can be
extracted without any hadronic uncertainty.
\end{abstract}  
  
\maketitle

\section{Introduction}
The theoretically cleanest way of determining the angle 
\begin{equation} 
\gamma=\arg(-V_{ud}V^*_{ub}/V_{cd}V_{cb}^*),
\end{equation}
is to utilize the interference between the $b\to c\bar{u}s$ and $b\to
u\bar{c} s$ decay amplitudes
\cite{Branco:1999fs,Gronau:1990ra,Gronau:1991dp,Dunietz,Atwood:1996ci,
Atwood:2000,Aleksan:2002mh,Soffer:1999dz,other,Grossman:2002aq,
Gronau:2002mu,Suprun:2003}.
Because these transitions involve only distinct quark flavors, there
are no penguin contributions to these decays. In the original idea by
Gronau and Wyler (GW) \cite{Gronau:1991dp} the $B^\pm\to D_{CP} K^\pm$
decay modes are used, where $D_{CP}$ represents a $D$ meson which
decays into a CP eigenstate.  The dependence on $\gamma$ arises from
the interference between the $B^\pm\to D^0 K^\pm$ and $B^\pm\to \DbarZ
K^\pm$ decay amplitudes. The main advantage of the GW method is that,
in principle, the hadronic parameters can be cleanly extracted from
data, by measuring the $B^\pm\to D^0 K^\pm$ and $B^\pm\to \DbarZ
K^\pm$ decay rates.

In practice, however, measuring $\gamma$ in this way is not an easy
task.  Due to the values of the CKM coefficients and color
suppression, the ratio between the two interfering amplitudes, $r_B$
[see Eq. (\ref{weakphase})], is expected to be small, of order
$10\%-20\%$. This reduces the sensitivity to $\gamma$, which is
roughly proportional to the magnitude of the smaller amplitude. In
addition, if the strong phases vanish, measuring $\gamma$ makes use
of terms of order $r_B^2$. In contrast, if a large strong phase is
involved in the interference, there is a sensitivity to $\gamma$ at
order $r_B$ with most methods. Thus, in general, having large 
interfering amplitudes
with large relative strong phases is a favorable situation.

Since the hadronic parameters are not yet known, it is still not clear
which of the proposed methods is more sensitive.  In addition, all the
methods are expected to be statistically limited.  It is therefore
important to make use of all modes and methods, as well as to try to find
new methods.  Any new method that is based on ``unused'' decay
channels increases the total statistics.  Moreover, many of the
analyses are sensitive to common hadronic parameters, for example,
$r_B$. Combining them will increase the sensitivity of the measurement
by more than just the increase in statistics.

Here we study the possibility to use $B^\pm\to D K^\pm$, followed by a
multibody $D$ decay, in order to cleanly determine $\gamma$. While
this idea was already discussed in \cite{Atwood:2000}, most of our
results and applications are new.  For the sake of concreteness, we
concentrate on the $D \to K_S\,\pi^-\pi^+$ decay mode.  The advantage
of using such decay chains is threefold. First, one expects large
strong phases due to the presence of resonances. Second, only Cabibbo
allowed $D$ decay modes are needed. Third, the final state involves
only charged particles, which have a higher reconstruction efficiency
and lower background than neutrals.  The price one has to pay is that
a Dalitz plot analysis of the data is needed.  We describe how to do
the Dalitz plot analysis in a model-independent way, and explore the
advantages gained by introducing verifiable model-dependence.  The
final balance between the advantages and disadvantages depends on
yet-to-be-determined hadronic parameters and experimental
considerations.

\section{Model independent determination of $\gamma$}\label{model-indep}
As we shall show in this section, to perform a model independent
determination of the angle $\gamma$ one needs to measure the two 
CP-conjugate decay
modes, $B^\pm\to D K^\pm\to (K_S\pi^-\pi^+)_D K^\pm$ and to perform a Dalitz
plot analysis of the $K_S\pi^-\pi^+$ final state originating from the
intermediate $D$ meson. (In the following discussion we neglect 
$D^0-\bar{D^0}$ mixing, which is a good approximation in the context
of the Standard Model. See appendix \ref{D-Dbar} for details.)

Let us first focus on the following cascade decay
\beqa
B^- \to D K^- \to (K_S \pi^- \pi^+)_D K^-,
\eeqa
and  define the amplitudes
\beqa
A(B^- \to D^0 K^-)&\equiv& A_B \label{AB},\\ 
A(B^- \to {\DbarZ} K^-) &\equiv& 
A_B r_B e^{i(\delta_B-\gamma)}.\label{weakphase}
\eeqa
The same definitions apply to the amplitudes for the CP conjugate
cascade $B^+ \to D K^+ \to (K_S \,\pi^+\pi^-)_D K^+$, with the change
of weak phase sign $\gamma\to -\gamma$ in \eqref{weakphase}. Since we
have set the strong phase of $A_B$ to zero by convention,
$\delta_B$ is the difference of strong phases between the two
amplitudes.  For the CKM elements, the usual convention of the weak
phases has been used. (The deviation of the weak phase from $-\gamma$
has been neglected, as it is suppressed by the factor $\lambda^4\sim 2
\times 10^{-3}$, with $\lambda$ being the sine of the Cabibbo angle.) 
The value of $|A_B|$ is known from the measurement of
the $B^-\to D^0 K^-$ decay width using flavor specific decays of $D^0$
and the precision of its determination is expected to further improve
\cite{experimentAB}. The amplitude $A(B^- \to \DbarZ K^-)$ is color
suppressed and cannot be determined from experiment in this way 
\cite{Atwood:1996ci}. The color suppression together
with the experimental values of the ratio of the relevant CKM
elements leads to the theoretical expectation $r_B\sim 0.1-0.2$ (see
recent discussion in \cite{Gronau:2002mu}).

For the three-body $D$ meson decay we define
\begin{equation}
\begin{split}\label{CP-for-D}
A_D(s_{12},s_{13}) \equiv A_{12,13}\,e^{i\delta_{12,13}} 
&\equiv  A(D^0 \to K_S(p_1) \pi^-(p_2) \pi^+(p_3))\\
& =A(\DbarZ \to K_S(p_1) \pi^+(p_2) \pi^-(p_3)),
\end{split}
\end{equation}
where $s_{ij}=(p_i+p_j)^2$, and $p_1,p_2,p_3$ are the momenta of the
$K_S, \pi^-,\pi^+$ respectively. We also set the magnitude
$A_{12,13}\ge0$, such that $\delta_{12,13}$ can vary between $0$ and
$2\pi$.  In the last equality the CP symmetry of the strong
interaction together with the fact that the final state is a spin zero
state has been used. With the above definitions, the amplitude for
the cascade decay is
\beqa
A(B^-\to (K_S \pi^- \pi^+)_D K^-)=A_B {\cal P}_D \big(A_D(s_{12},s_{13}) + 
r_B e^{i(\delta_B-\gamma)}A_D(s_{13},s_{12})\big), \label{amplitude}
\eeqa
where  ${\cal P}_D$ is the $D$ meson propagator.  Next, we write down the
expression for the reduced partial decay width
\begin{equation}
\begin{split}
 d\hat\Gamma(B^- \to (K_S \pi^-\pi^+)_D K^-)=
 \Big(&A_{12,13}^2 +r_B^2 \, A_{13,12}^2 \\ &
+ 2 r_B
\Re\left[A_D(s_{12},s_{13})\,A_D^*(s_{13},s_{12})\,e^{-i(\delta_B-\gamma)}
\right]\Big) dp ,
\end{split}\label{decay-width}
\end{equation}
where $dp$ denotes the phase space variables, and we used 
the extremely accurate  narrow width
approximation for the $D$ meson propagator.

In general, there is no symmetry between the two arguments of $A_D$ in
Eq. \eqref{amplitude}, and thus in the rates over the Dalitz plot. A
symmetry would be present if, for instance, the three-body $D$ decay
proceeded only through $\rho$-like resonances.  We emphasize,
however, that the product $A_D(s_{12},s_{13})\,A_D^*(s_{13},s_{12})$
in the interference term in Eq. \eqref{decay-width} is symmetric under
the exchange $s_{12} \leftrightarrow s_{13}$ followed by complex
conjugation. This fact is used to simplify the analysis.

The moduli of the $D$ decay amplitude $A_{12,13}$ can be measured
from the Dalitz plot of the $D^0\to K_S\pi^-\pi^+$ decay. To perform
this measurement the flavor of the decaying neutral $D$ meson has to
be tagged. This can be best achieved by using the charge of the soft
pion in the decay $D^{*+}\to D^0 \pi^+$. 
However, the phase $\delta_{12,13}$ of the $D$ meson decay amplitude 
is not measurable without further model dependent
assumptions. The cosine of the relevant phase difference may
be measured at a charm factory (see section~\ref{improving}). 
If the three-body decay $D^0\to K_S\pi^-\pi^+$ is assumed to
be resonance dominated, the Dalitz plot can be fit to a sum of
Breit-Wigner functions, determining also the relative phases of the
resonant amplitudes.  This is further discussed in section
\ref{Breit-Wigner}.  Here we assume that no  charm factory data is available
and develop the formalism without any model
dependent assumptions.

Using the trigonometric relation 
$\cos(a+b)=\cos a\, \cos b - \sin a \, \sin b$, 
the last term of \eqref{decay-width} can be written as
\beqa
&&
\Re\left[A_D(s_{12},s_{13})\,A_D^*(s_{13},s_{12})\,e^{-i(\delta_B-\gamma)}
\right]=\\
&& A_{12,13}\,A_{13,12}\left[
\cos(\delta_{12,13}-\delta_{13,12})\cos(\delta_B-\gamma)+
\sin(\delta_{12,13}-\delta_{13,12})\sin(\delta_B-\gamma)\right]. \nonumber
\eeqa
Obviously, to compare with the data, an
integration over at least some part of the Dalitz plot has to be
performed. We therefore partition the Dalitz plot into $n$ bins and
define
\begin{subequations}\label{defvar}
\begin{align}
c_i &\equiv \int_i dp\; 
A_{12,13}\,A_{13,12}\,\cos(\delta_{12,13}-\delta_{13,12}),
\label{ci}\\
s_i &\equiv \int_i dp\;  
A_{12,13}\,A_{13,12}\sin(\delta_{12,13}-\delta_{13,12}),
\label{si}\\
T_i &\equiv \int_i dp\;  
A_{12,13}^2, \label{Bi}
\end{align}
\end{subequations}
where the integrals are done over the phase space of the $i$-th
bin. The variables $c_i$ and $s_i$ contain differences of strong phases and
are therefore unknowns in the analysis. The variables $T_i$, on
the other hand, can be measured from the flavor tagged $D$ decays as
discussed above, and are assumed to be known inputs into the analysis.

\begin{figure}
\begin{center}
\epsfig{file=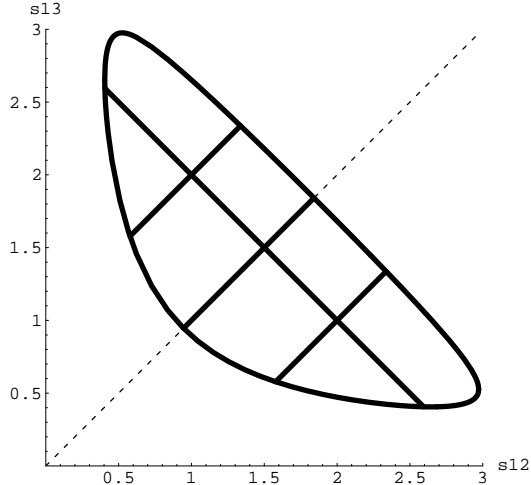, height=7.0cm}
\caption{\footnotesize{The partitions of Dalitz plot as discussed in
text. The symmetry axis is the dashed line. On the axes we have $s_{12}=m_{K_s\pi^-}^2$ and $s_{13}=m_{K_s\pi^+}^2$ in $\text{GeV}^2$. }} \label{fig1}
\end{center}
\end{figure}

Due to the symmetry of the interference term, it is convenient to use
pairs of bins that are placed symmetrically 
about the $12 \leftrightarrow 13$ line, as
shown in Fig. \ref{fig1}. Consider an even, $n=2k$, number of bins. The
$k$ bins lying below the symmetry axis are denoted by index $i$,
while the remaining bins  are indexed with $\bar{i}$.
The $\bar{i}$-th bin is obtained by  mirroring the $i$-th bin over
the axis of symmetry. The variables
$c_i, s_i$ of the $i$-th bin are related to the variables of the
$\bar{i}$-th bin by
\begin{equation}
c_{\,\bar{i}}=c_i, \qquad s_{\,\bar{i}}=-s_i, \label{vanish}
\end{equation}
while there is no relation between $T_i$ and $T_{\bar{i}}$.
Note that had one used $12\leftrightarrow 13$ symmetric bins centered
on the symmetry axis, one would have had  $s_i=0$.

Together with the information available from the $B^+$ decay, we 
arrive at a set of $4k$ equations
\begin{subequations}\label{relations4k}
\begin{align}
\begin{split}\label{11a}
\hat\Gamma^-_i \equiv 
\int_i d\hat\Gamma(B^- \to (K_S \pi^-\pi^+)_D K^-)&=\\
T_i  +r_B^2 T_{\bar{i}}\; +& 
2 r_B [\cos(\delta_B-\gamma) c_i + \sin(\delta_B-\gamma) s_i],
\end{split}
\\
\begin{split}
\hat\Gamma^-_{\bar{i}} \equiv 
\int_{\bar{i}} d\hat\Gamma(B^- \to (K_S \pi^- \pi^+)_D K^-)&=\\
T_{\bar{i}} +r_B^2 T_i \;+ &
2 r_B [\cos(\delta_B-\gamma) c_i - \sin(\delta_B-\gamma) s_i],
\end{split}
\\
\begin{split}
\hat\Gamma^+_i \equiv\int_i d\hat\Gamma(B^+ \to (K_S \pi^- \pi^+)_D K^+)&=\\
T_{\bar{i}} +r_B^2 T_i \;+ &
2 r_B [\cos(\delta_B+\gamma) c_i - \sin(\delta_B+\gamma) s_i],
\end{split}
\\
\begin{split}
\hat\Gamma^+_{\bar{i}} \equiv 
\int_{\bar{i}} d\hat\Gamma(B^+ \to (K_S \pi^- \pi^+)_D K^+)&=\\
T_i +r_B^2 T_{\bar{i}} \;+ &
2 r_B [\cos(\delta_B+\gamma) c_i + \sin(\delta_B+\gamma) s_i].
\end{split}
\end{align}
\end{subequations}
These equations are related to each other through
$12\leftrightarrow 13$ and/or $\gamma \leftrightarrow -\gamma$  exchanges.
All in all,
there are $2k + 3$ unknowns in \eqref{relations4k},
\beqa \label{totpar}
c_i, ~s_i, ~r_B, ~\delta_B, ~\gamma,
\eeqa
so that the $4k$ relations \eqref{relations4k} are solvable for $k\ge
2$. In other words, a partition of the $D$ meson Dalitz plot to four
or more bins allows for the determination of $\gamma$
without hadronic uncertainties.  This is our main result.

Alternatively to this binning, 
one can use a partition of the Dalitz plot into $k$ bins
which are symmetric under $12\leftrightarrow 13$. For that case,
$s_i=0$ and the set of the $4k$ equations \eqref{relations4k} reduces
to $2k$ relations (the first two and the last two equations in
\eqref{relations4k} are the same in this case). Then, there are just
$k+3$ unknowns to be solved for, which is possible for $k\ge 3$.
While such binning may be needed due to low statistics, it 
has several disadvantages, which are further discussed below.

When $c_i=0$ or $s_i=0$ for all $i$, some equations become degenerate
and $\gamma$ cannot be extracted. However, due to resonances, we do
not expect this to be the case. Degeneracy also occurs if
$\delta_B=0$.  In this case, $\gamma$ can still be extracted if some
of the $c_i$ and/or $s_i$ are independently measured, as discussed in the
following sections.

The optimal partition of the Dalitz plot as well as the number of bins
is to be determined once the analysis will be done. Some of the
considerations that enter this choice are as
follows.  First, one would like to have as many small bins as
possible, in order that $c_i$ and $s_i$ do not average out to small
numbers.  Second, the bins have to be large enough that there are
significantly more events than bins. Otherwise there will be more
unknowns than observables. There are also
experimental considerations, such as optimal parameterization of 
backgrounds and reconstruction efficiency.

\section{Improved Measurement of \boldmath $c_i$ and $s_i$}
\label{improving}

So far, we have used the $B$ decay sample to obtain all the unknowns,
including $c_i$ and $s_i$, which are parameters of the charm system.
We now discuss ways to make use of high-statistics charm decays to
improve the measurement of these parameters, or obtain them
independently. Doing so will reduce the number of unknowns
that need to be determined from the relatively low-statistics $B$
sample, thereby reducing the error in the measurement of $\gamma$.

The first improvement in the measurement is obtained by making use of
the large sample of tagged $D$ decays, identified in the decay
$D^{*+}\to D^0 \pi^+$, at the $B$ factories. So far we only assumed
that we use this data to determine $T_i$. In fact, it can also be used
to bound the  unknowns $c_i$ and $s_i$ defined in \eqref{defvar}:
\begin{align}
|s_i|,\;|c_i|& \le \int_i dp\; A_{12,13} A_{13,12}  
\le \sqrt{T_i \,T_{\bar{i}}}\,.
\end{align}
This bound will help decrease the error in the determination of $\gamma$,
with an especially significant effect when, due to low statistics in
each bin, $c_i$ and $s_i$ will be determined with large errors.

Next, we show that the $c_i$ can be independently measured at a charm
factory~\cite{Soffer:1998un,Silva:1999bd,Gronau:2001nr}. This is done
by running the machine at the $\psi(3770)$ resonance, which decays
into a $D\Dbar$ pair. If one $D$ meson is detected in a CP eigenstate
decay mode, it tags the other $D$ as an eigenstate of the opposite CP
eigenvalue. The
amplitude and partial decay width for this state to decay into the
final state of interest are
\beqa
A(D^0_\pm \to K_S(p_1) \pi^-(p_2) \pi^+(p_3)) &=& \frac{1}{\sqrt{2}} \left(
	A_D(s_{12},s_{13}) \pm A_D(s_{13},s_{12}) \right), 
	\\
d\Gamma(D^0_\pm \to K_S(p_1) \pi^-(p_2) \pi^+(p_3)) &=&
	\frac{1}{2} \left(
	A_{12,13}^2 + A_{13,12}^2 \right)
	\pm A_{12,13} A_{13,12} 
		\cos(\delta_{12,13}-\delta_{13,12})dp.~~\nonumber
\eeqa
where we defined $D^0_\pm \equiv (D^0 \pm \DbarZ)/\sqrt{2}$.
With these relations, one readily obtains 
\begin{equation}
c_i =  
	\frac{1}{2} \left[\int_i 
	 d\Gamma(D^0_+ \to K_S(p_1) \pi^-(p_2) \pi^+(p_3))
	-\int_id\Gamma(D^0_- \to K_S(p_1) \pi^-(p_2) \pi^+(p_3))
	\right].
\label{ci-charm-factory}
\end{equation}
As stated above, obtaining this independent measurements reduces the 
error in the measurement of $\gamma$ by removing $k$ of the $2k+3$ unknowns.

We can further improve the measurement if we take each bin $i$ and
further divide it into $n_i$ sub-bins, such that the quantities
$A_{12,13}$, 
$\cos(\delta_{12,13}-\delta_{13,12})$,
and 
$\sin(\delta_{12,13}-\delta_{13,12})$
do not change significantly within each sub-bin $i'$. Naively, this
statement appears to introduce model dependence. In practice, however,
the high statistics in the tagged $D$ sample and the charm factory
$\psi(3770)$ sample allow its verification up to a statistical error,
which can be measured and propagated to the final measurement of $\gamma$.

Given this condition, Eq.~(\ref{ci}) may be written as 
\begin{eqnarray}
c_i =\sum_{i'} c_{i'}&=& \sum_{i'}
A_{i'}\,A_{\overline{i'}}\cos(\delta_{i'}-\delta_{\overline{i'}})
\Delta p_{i'}
	= \sum_{i'}\sqrt{T_{i'} T_{\overline{i'}}} \;
		\cos(\delta_{i'}-\delta_{\overline{i'}}),
\label{ci-improved}
\end{eqnarray}
where 
the $\overline{i'}$-th sub-bin is the $12\leftrightarrow 13$
mirror image of the $i'$-th sub-bin, 
$A_{i'}$ and $\delta_{i'}$ are the values of $A_{12,13}$ and
$\delta_{12,13}$ on sub-bin $i'$, taken to be constant throughout the sub-bin,
and $\Delta p_{i'}$ is the area of sub-bin $i'$. 
Analogously to Eq.~(\ref{Bi}), we have defined the quantities
$T_{i'} = A_{12,13}^2 \Delta p_{i'}$, which are measured using
the tagged $D$ sample.
The $c_{i'}$'s are assumed to be measured at the charm
factory, applying \eqref{ci-charm-factory} to the sub-bin $i'$. 
Similarly, Eq.~(\ref{si}) becomes
\begin{eqnarray}
s_i = \sum_{i'} \sqrt{T_{i'} T_{\overline{i'}}} \;
		\sin(\delta_{i'}-\delta_{\overline{i'}}) 
	= \sum_{i'} \pm \sqrt{T_{i'} T_{\overline{i'}} -c_{i'}^2} 
		.\label{si-improved}
\end{eqnarray}
Eq.~(\ref{si-improved}) removes the $k$
unknowns $s_i$, and replaces them with the two-fold ambiguity
associated with the sign of the square root. Thus, the best approach
is to have the signs of $s_i$ determined by the fit, while constraining
their absolute values to satisfy Eq.~(\ref{si-improved}).
Doing so will reduce the ``strain'' on the $B$ decay sample,
reducing the error on $\gamma$.

Another option for removing the dependence on $s_i$ is to use bins
centered symmetrically about the $12\leftrightarrow 13$ line, making
$s_i$ vanish, as discussed after Eq. \eqref{vanish}. In this case,
both the number of unknowns and the number of observables (bins) is
reduced by $k$. By contrast, using Eq.~(\ref{si-improved}) introduces
new information from the independent tagged $D$ sample, and is
therefore preferred. Doing so also preserves the $\sin(\delta_B -
\gamma)$ terms in Eq.~(\ref{relations4k}), which helps resolve
discrete ambiguities (see~\cite{Soffer:1999dz} and section~\ref{discussion}).

\section{Assuming Breit-Wigner dependence}\label{Breit-Wigner}
If the functional dependence of both the moduli and the phases of the
$D^0$ meson decay amplitudes $A_D(s_{12},s_{13})$ were known, then the
analysis would be simplified.  There would be only three
variables, $r_B, \delta_B$, and $\gamma$, that need to be fit to the
reduced partial decay widths in Eq.~\eqref{decay-width}. A plausible
assumption about their forms, which is also supported by experimental
data
\cite{Frabetti:1994di,Albrecht:1993jn,Muramatsu:2002jp}, is that a
significant part of the three-body $D^0\to K_S \pi^- \pi^+$ decay
proceeds via resonances. These include decay transitions of the form
$D^0\to K_S \rho^0 \to K_S \pi^- \pi^+$ or $D^0\to K^{*-}(892) \pi^+\to
K_S \pi^- \pi^+$, as well as decays through higher resonances,
e.g., $f_0(980)$, $f_2(1270)$, or $f_0(1370)$, inducing $\rho$-like
transitions, or $K^*_0(1430)$, which induces a $K^*(892)$-like
transition.

It is important to stress that these assumptions can be tested. By
making use of the high statistics tagged $D$ sample, one can test
that the assumed shapes of the resonances are consistent with the
data. While the error introduced by using the Breit-Wigner shapes is
theoretical, it is expected to be much smaller than the statistical
error in the measurement of $\gamma$.  It will become a problem only
when the $B$ sample is large enough to provide a precision measurement
of $\gamma$.  By then the tagged $D$ sample will have increased as
well, allowing even more precise tests of these assumptions, as well
as improving the precision of the methods presented in
section~\ref{improving}.

The decay amplitude can then be fit to a sum of Breit-Wigner
functions and a constant term. Following the notations of Ref.
\cite{Aitala:2001zx} we write
\begin{equation} 
\begin{split}
A_D(s_{12},s_{13})&=A(D^0 \to K_S(p_1) \pi^-(p_2) \pi^+(p_3))=\\ &=
a_0 e^{i\delta_0}+\sum_r a_r e^{i\delta_r} {\cal A}_r 
(s_{12}, s_{13}),
\end{split}\label{resonansatz}
\end{equation}
where the first term corresponds to the non-resonant term and the second
to the resonant contributions. The Breit-Wigner function is defined as
\beqa \label{defBW}
{\cal A}_r  (s_{12}, s_{13}) = {^J\!\!{\cal M}_r} \times {BW}^r,
\eeqa
where $r$ represent a specific resonance in either the
$K_S(p_1) \pi^-(p_2)$,  $K_S(p_1)\pi^+(p_3)$ or $\pi^-(p_2)
\pi^+(p_3)$ channel.
${^J\!\!{\cal M}_r}$ is the term which accounts for the angular
dependence. It depends on the spin $J$ of the resonance.  For
example, ${^0\!{\cal M}_r=1}$ and ${^1\!{\cal M}_r}= -2 \vec{k}_1\cdot
\vec{k}_3$. Here $\vec{k}_1,\vec{k}_3$ are, respectively, the three
momenta of one of the particles originating from the resonance and of
the remaining particle, as measured in the rest frame of the two
resonating particles \cite{Aitala:2001zx}.  ${BW}^r$ corresponds to
the relativistic Breit-Wigner function and is given by
\begin{equation}\label{Bwr}
BW^r(s)=\frac{1}{s-M_r^2+ iM_r \Gamma_r(\sqrt{s})},
\end{equation}
where $M_r$ is the mass of the $r$-th resonance and $\Gamma_r(\sqrt{s})$ 
denotes the mass-dependent width. The argument of $BW^r$ is $s_{12}$
[$s_{13}$,  $s_{23}$]
for a $K_S(p_1) \pi^-(p_2)$ [$K_S(p_1) \pi^+(p_3)$, $\pi^-(p_2)
\pi^+(p_3)$] resonance.
One can find detailed expressions for all the functions mentioned above in 
Ref. \cite{Aitala:2001zx}.

One of the strong phases $\delta_i$ in the ansatz \eqref{resonansatz} 
can be put to zero, while others are fit to the experimental data
together with the amplitudes $a_i$.
The best option is to fit 
the Dalitz plot of tagged $D$ decays, as was done a decade
ago by the ARGUS and E687 collaborations
\cite{Frabetti:1994di,Albrecht:1993jn} and recently by the CLEO
collaboration \cite{Muramatsu:2002jp}. The obtained functional form of
$A_D(s_{12},s_{13})$ can then be fed to Eq. \eqref{decay-width}, which
is then fit to the Dalitz plot of the $B^\pm\to (K_S \pi^- \pi^+)_D
K^\pm$ decay with $r_B$, $\delta_B$ and $\gamma$ left as free
parameters. In appendix
\ref{app:BW} we provide a formula for the latter case, where only
three resonance are included in the analysis.

\section{Discussions}\label{discussion}

The observables $\hat\Gamma^\pm_i$ defined in 
\eqref{relations4k} can be used to experimentally look for
direct CP violation. Explicitly,
\beqa \label{dircp}
a_{\rm CP}^i&\equiv& \hat\Gamma^-_i - \hat\Gamma^+_{\bar{i}} = 
4 r_B \sin\gamma \left[c_i \sin \delta_B - s_i  \cos \delta_B\right] ,
\nonumber \\
a_{\rm CP}^{\bar{i}}&\equiv& \hat\Gamma^-_{\bar i} - \hat\Gamma^+_i = 
4 r_B \sin\gamma \left[c_i \sin \delta_B + s_i  \cos \delta_B\right] .
\eeqa 
It is manifest that finite $a_{\rm CP}$ requires non vanishing strong
and weak phases. The first terms in the parenthesis in \eqref{dircp}
depends on $\sin\delta_B$. This is the same dependence as for a two-body 
$D$ decays into CP eigenstates. In the second terms, which depend
on $\cos\delta_B$, the required strong phase arises from the $D$ decay
amplitudes. Due to the resonances, we expect this strong phase to be
large. Therefore, it may be that direct CP
violation can be established in this mode even before the full
analysis to measure $\gamma$ is conducted. With more data, $\gamma$ can be
extracted assuming Breit-Wigner resonances (cf. section
\ref{Breit-Wigner}).  Eventually, a model independent extraction of
$\gamma$ can be done (cf. section~\ref{model-indep} and~\ref{improving}).

The above proposed method for the model independent measurement of
$\gamma$ involves a four-fold ambiguity in the extracted value. The
set of equations
\eqref{relations4k} are invariant under each of the two discrete
transformations
\beqa
P_\pi\equiv \{ \delta_B\to \delta_B + \pi, \gamma\to \gamma +\pi\},
\qquad
P_-\equiv \{ \delta_B\to -\delta_B, \gamma\to -\gamma,s_i\to -s_i\}. 
\eeqa
We note that if all the bins used are symmetric under
$12\leftrightarrow 13$, the absence of the $\sin(\delta_B -
\gamma)$ terms in Eq.~(\ref{relations4k}) introduces a new ambiguity
transformation, $P_{\rm ex} \equiv \gamma \to \delta_B, \delta_B \to \gamma$.
The discrete transformation $P_\pi$ is a symmetry of the amplitude
\eqref{amplitude} and is thus an irreducible uncertainty of the
method. It can be lifted if the sign of either $\cos\delta_B$ or
$\sin\delta_B$  is known. The ambiguity due to $P_-$ can be resolved if
the sign of $\sin\delta_B$ is known or if the sign of $s_i$ can be
determined in at least some part of the Dalitz plot. The latter can be
done by fitting a part of the Dalitz plot to Breit-Wigner
functions. We emphasize that only the sign of the phase of the
resonance amplitude is required, and thus we can safely use a
Breit-Wigner form for this purpose.

The $r_B$ suppression present in the scheme outlined above can be
somewhat lifted if the cascade decay $B^-\to D X_s^- \to (K_S \pi^-
\pi^+)_D X_s^-$ is used \cite{Gronau:2002mu,Aleksan:2002mh}.  Here
$X_s^-$ is a multibody hadronic state with an odd number of kaons
(examples of such modes are $K^-\pi^-\pi^+$, $K^-\pi^0$ and $K_S \pi^-
\pi^0$). Unlike the $B^-\to \DbarZ K^-$ decay, these modes have
color-allowed contributions.  This lifts the color suppression in
$r_B$, while the mild suppression due to the CKM matrix elements
remains. The major difference compared with the case of the two-body $B^-$
decay is that now $r_B$ and $\delta_B$ are functions of the $B\to D
X_s^-$ decay phase space. Therefore, the experimental analysis has to
deal with two Dalitz plots, one describing $B\to D X_s^-$ and the
other describing the $D\to K_S \pi^-\pi^+$ decay. In
appendix \ref{app:multi} the necessary formalism that applies to this
case is outlined.  Note that the above mentioned treatment for
multibody $B$ decays also applies to quasi two-body $B$ decays involving a
resonance, such as $B\to D K^*$.

In addition to using different $B$ modes, statistics may be increased by 
employing various $D$ decay modes as well. An interesting possibility
is the Cabibbo allowed $D\to K_S
\pi^-\pi^+ \pi^0$ decay. It comes with an even larger branching ratio
than the $D\to K_S \pi^-\pi^+$ decay. In addition, it has many intermediate
resonances contributing to the greatly varying decay amplitude, which
is what is needed for the extraction of $\gamma$. 
The disadvantages of this mode are the low reconstruction efficiency
of the $\pi^0$, as well as the binning difficulties introduced by the
higher dimensionality of the four-body phase space.
The formalism of
section \ref{model-indep} applies to this mode as well, but now the
partition of the four-body phase space is meant in
Eq. \eqref{relations4k}. In the 
equivalent of \eqref{CP-for-D}, this mode has an extra minus sign, since we
have introduced a new CP-odd state, the $\pi^0$. The final set of equations is
then obtained from \eqref{relations4k} by replacing $r_B\to -r_B$.
The Cabibbo allowed 
mode $D \to K^- K^+ K_S$ may also be used for the extraction
of $\gamma$, as can the Cabibbo
suppressed decays to $K^- K^+ \pi^0$, $\pi^- \pi^+ \pi^0$,
and $K_S K^+ \pi^-$. One can also use (almost) flavor eigenstate
decay modes, such as $D \to K^- \pi^+ \pi^0$ and $D \to K^- \pi^+
\pi^- \pi^+$ \cite{Atwood:2000}. 
Here, the important interference is between the Cabibbo allowed
$\Dbar$ decay and the doubly Cabibbo suppressed $D$ decay.

While we concentrated on charged $B$ decays, the Dalitz plot analysis
presented here can also be applied to self-tagging decays of neutral $B$
mesons~\cite{Dunietz}. It is also straightforward to apply it to
cases where time dependent CP asymmetries are measured \cite{Gronau:1990ra}.

The sensitivity to $\gamma$ is roughly proportional to the smaller of
the two interfering amplitudes.  Assuming that the only two small
parameters are $r_B$ and $\lambda$, our method is sensitive to
$\gamma$ at $O(r_B)$. However, the method is sensitive to $\gamma$
only in parts of the Dalitz plot.  The highest sensitivity is in
regions with two or more overlapping resonances.  The sensitivity of
the proposed method is therefore of order $O(r_B \xi)$, where $\xi^2$
is the fraction of events which are in the interesting region of the
Dalitz plot.

A crucial point of our method is that it uses interference between two
Cabibbo allowed $D$ decay amplitudes. This is against the common
intuition, which suggests that we must have a $\lambda^2$ suppression
for such interference to take place, as we need a final state that is 
common to both
$D$ and $\Dbar$. Specifically, one typically requires one Cabibbo
allowed decay and another that is doubly Cabibbo suppressed, or two
decays that are singly Cabibbo suppressed. To overcome this
preconception, our method makes use of $K^0-\overline K^0$ mixing
(which is also the case for the two-body $D\to K_S \pi^0$ decay), plus
the existence of overlapping resonances, which are obtained by
Cabibbo-allowed $D^0$ and $\DbarZ$ decays.  In addition, it is
important that the hadronic three-body $D$ meson decays have a widely
changing amplitude over the Dalitz plot, which is ensured by the presence of
resonances in this energy region. If the strong phases
$\delta_{12,13}$ and the moduli $A_{12,13}$ in Eq. \eqref{defvar} were
(almost) constant across the available phase space, the extraction of
$\gamma$ from Eqs. \eqref{relations4k} would not be possible.

Before concluding, we mention that quasi two-body $D$ decays where one
of the particles is a resonance, 
such as $D\to K^{*+}\pi^- $ and $ D\to
K^{+}\rho^-$~\cite{Atwood:1996ci},
were proposed for use in measuring
$\gamma$. But in fact, using 
such decays requires a Dalitz plot analysis (see
e.g. \cite{Grossman:2002aq,Suprun:2003}). What we showed here is that
one can actually use the whole Dalitz plot to carry out the analysis
and does not need to single out contributions of one particular
resonance. Moreover, we showed that  the
assumption about the shapes of the resonances can be avoided,
essentially with currently available data-sets.

In conclusion, we have shown that the angle $\gamma$ can be determined
from the cascade decays $B^\pm\to K^\pm (K_S \pi^- \pi^+)_D$.  The
reason for the applicability of the proposed method lies in the
presence of resonances in the three-body $D$ meson decays that provide
a necessary variation of both the phase and the magnitude of the decay
amplitude across the phase space. The fact that no Cabibbo suppressed
$D$ decay amplitudes are used in the analysis is another advantage of
the method.  However, it does involve a Dalitz plot analysis with
possibly only parts of the Dalitz plot being practically useful for
the extraction of $\gamma$. In reality, many methods have to
be combined in order to achieve the required statistics for a precise
determination of $\gamma$ \cite{Soffer:1999dz}.

\begin{acknowledgments}
We thank Michael Gronau, Zoltan Ligeti and Marie-Helene Schune for
helpful discussions.  Y.G. is supported in part by the Israel Science
Foundation (ISF) under Grant No.~237/01, by the United States--Israel
Binational Science Foundation (BSF) through Grant No.~2000133 and by
the German--Israeli Foundation for Scientific Research (GIF) through
Grant No. G-698-22.7/01.  The work of A.S. was supported by the
U.S. Department of Energy under contract DE-FG03-93ER40788.  J.Z. is
supported in part by the Ministry of Education, Science and Sport of
the Republic of Slovenia.
\end{acknowledgments}

\appendix
\section{The effect of $D-\overline{D}$ mixing}\label{D-Dbar}
In this section we 
focus on the contributions introduced by the fact that the flavor states
$|D^0\rangle$, $|\DbarZ\rangle$ and the mass eigenstates
$|D_{H,L}\rangle=p_D |D^0\rangle\pm q_D|\DbarZ\rangle$ do not
coincide. This effect was studied in the general case in
Ref. \cite{Amorim:1998pi}. Here we apply their formalism to our case.

Following Ref. \cite{Amorim:1998pi} we
introduce the rephasing-invariant parameter $\chi_1$ 
\beqa\label{chi}
\chi_1=\frac{\lambda_{D\to f}+\xi_{B^-\to D}}{1+\lambda_{D\to f} 
\xi_{B^-\to D}},
\eeqa 
where
\beqa
\lambda_{D\to f}=\frac{q_D}{p_D} \frac{A_{\DbarZ \to f}}{A_{D^0\to
f}}, \qquad
\xi_{B^-\to D}=\frac{A_{B^-\to \DbarZ K^-}}{A_{B^-\to D^0 K^-}}
\frac{p_D}{q_D}=r_B e^{-i(2 \theta_D-\delta_B +\gamma)},
\eeqa
and we use the definitions of Eqs. \eqref{AB} and \eqref{weakphase}
and allow for new physics effects in $q_D/p_D=e^{i2 \theta_D}$. (In the phase
convention where the $D$ decay amplitudes are real, the phase $\theta_D$
is negligible in the Standard Model).  In our case, the final state $f$
equals $K_S \pi^- \pi^+$, which leads to
\beqa
\lambda_{D\to K_S(p_1) \pi^-(p_2)\pi^+(p_3)}=
e^{i 2\theta_D}\frac{A_D(s_{13},s_{12})}
{A_D(s_{12},s_{13})}=R_D(s_{12,},s_{13})
e^{i (2 \theta_D+\delta_{13,12}-\delta_{12,13})}.
\eeqa
Once $D-\Dbar$ mixing is taken into account in the
analysis, the expression for the partial decay width
\eqref{decay-width} is multiplied by the correction term
\cite{Amorim:1998pi}
\beqa\label{corrections}
1-\Re(\chi_1) y_D+\Im(\chi_1)x_D,
\eeqa
where we have expanded the correction term to first
order in the small parameters
\beqa
x_D=\frac{\Delta m}{\Gamma}, \qquad
y_D=\frac{\Delta \Gamma}{2\Gamma},
\eeqa
where $\Delta m$ and $\Delta \Gamma$ are the mass and decay width differences
in the $D-\Bar{D}$ system, and $\Gamma$ is the $D^0$ decay width.  The
values of $x_D$ and $y_D$ are constrained by present measurements to
be in the percent range, $y_D=(1.0\pm 0.7)\%$
\cite{Grothe:2003zg} and $|x|<2.8 \%$ \cite{Godang:1999yd} (assuming
small strong phases).

The ratio of magnitudes, $R_D(s_{12,},s_{13})$, depends on the
position in the Dalitz plot, and can vary widely.  Our method is
useful for the model independent extraction of $\gamma$ only in the
region where $R_D$ is of order one. We therefore distinguish three
limiting cases
\begin{itemize}
\item
$R_D\gg 1 \gg r_B$, for which  $\Re(\chi_1),
\Im(\chi_1)\sim O(1/r_B)$ and therefore the corrections in
\eqref{corrections} can be of order $10\%$. However, this is the
region of Dalitz plot where our method is mostly not sensitive to $\gamma$
and therefore the induced corrections due to $D-\bar{D}$ mixing do not
translate into an error on the extracted $\gamma$.
\item
$R_D\sim 1 \gg r_B$, for which $\Re(\chi_1), \Im(\chi_1)\sim O(1)$ and
therefore the corrections in \eqref{corrections} are at the percent
level. This is the value of $R_D$ for which our method is most sensitive to
$\gamma$.
\item
$ 1 \gg r_B\sim R_D$, for which $\Re(\chi_1), \Im(\chi_1)\sim
O(r_B,R_D)$ and therefore the corrections in \eqref{corrections} are
very small.
\end{itemize}
In conclusion, we expect errors of at most a few percent due to
neglecting $D-\bar{D}$ mixing in our method.  In principle, even
these errors can be taken into account
\cite{Meca:1998ee,Amorim:1998pi,Silva:1999bd}.

\section{A fit to Breit-Wigner functions: an illustration for three resonances}
\label{app:BW}
In this appendix we provide the formulae for the fit of $D$ meson
decay amplitude to a sum of three Breit-Wigner functions describing
$K^{*\pm}(892)$ and $\rho^0$ resonances.
We write Eq. \eqref{resonansatz} explicitly as
\begin{equation}
\begin{split}
A_D(s_{12},s_{13})&=A(D^0 \to K_S(p_1) \pi^-(p_2) \pi^+(p_3))=\\ &=
\Arhok\bwtt (s_{23}) + 
\Akpi e^{i \delta_F}\bwotw (s_{12}) + 
\Akpi r_D e^{i \delta_D} \bwoth (s_{13}) , 
\end{split}\label{resonansatz-exp}
\end{equation}
where $\delta_F$ ($\delta_D$) is the strong phase of the Cabibbo
favored (doubly Cabibbo suppressed) $D^0\to K^{*-}\pi^+$ ($D^0\to
K^{*+}\pi^-$) decay with respect to the decay $D^0\to K_S\rho^0$. We
further introduced
\beqa
\Arhok &\propto& A(D^0 \to \rho^0 K_S) = A(\DbarZ \to \rho^0 K_S),  \nonumber
\\
\Akpi e^{i \delta_F}&\propto& A(D^0 \to K^{*-} \pi^+) =  
A(\DbarZ \to K^{*+} \pi^-),
\nonumber \\
\Akpi r_D e^{i \delta_D} &\propto& A(D^0 \to K^{*+} \pi^-) =  
A(\DbarZ \to K^{*-} \pi^+).
\eeqa 
The Breit-Wigner functions ${\cal A}_r$ are defined in \eqref{defBW},
where we write in \eqref{resonansatz-exp} only the $s_{ab}$ dependence
of the $BW^r$ part, given in \eqref{Bwr}. The first index of $s_{ab}$ is
understood to denote also the particle appearing in the expression for
$^1\!{\cal M}_r$ \eqref{defBW}. Exchanging $a\leftrightarrow b$ corresponds to
$^1\!{\cal M}_r\leftrightarrow -^1\!{\cal M}_r$, in particular ${\cal
A}_{\rho^0}(s_{23})=-{\cal A}_{\rho^0}(s_{32})$.
In the above we assumed that there is no CP violation in the $D$
decays amplitudes. Note that there are two small parameters
\beqa
r_B \sim 0.1-0.2, \qquad
r_D \sim \lambda^2 \sim 0.05.
\eeqa
We then obtain (cf. \eqref{amplitude})
\beqa
&& A(B^-\to (K_S(p_1) \pi^-(p_2) \pi^+(p_3))_D K^-) = \\
&&~~~~
A_B {\cal P}_D \times \Big\{\left(\Arhok \bwtt(s_{23})+
 \Akpi  
\left[e^{i \delta_F} \bwotw(s_{12})+
r_D e^{i \delta_D} \bwoth(s_{13})\right]\right) +
\nonumber \\
&&~~~~ ~~~~~~~~~~~~r_B e^{i(\delta_B-\gamma)}\left(\Arhok \bwtt(s_{32})+
 \Akpi
 \left[e^{i \delta_F} \bwoth(s_{13})+r_D e^{i \delta_D}
\bwotw(s_{12})\right]\right)
\Big\}. \nonumber
\eeqa
The corresponding expressions for $B^+$ decays are obtained 
by changing $\gamma \to -\gamma$ and  $\pi^-(p_2) \pi^+(p_3) \to
\pi^+(p_2) \pi^-(p_3)$.

We further define
\beqa
\delta_-= \arg[\bwotw(s_{12})], \quad
\delta_+ = \arg[\bwoth(s_{13})], \quad
\delta_0 = \arg[\bwtt(s_{23})].~~~~
\eeqa
where the dependence of $\delta_{\pm, 0}$ on the position in the Dalitz
plot is implicitly assumed.
The reduced differential decay rate
is then
\beqa \label{master}
&& \!\!\!\!\!\!\!\!\!\! d\hat\Gamma(B^- \to (K_S \pi^- \pi^+)_D K^-)
\propto \nonumber \\ &&
\Arhok^2|\bwtt(s_{23})|^2 \left(1-2 r_B
  \cos(\delta_B-\gamma)+r_B^2\right)  +\nonumber \\ &&
\Akpi^2|\bwotw(s_{12})|^2 \left(1+2 r_B r_D
  \cos(\delta_{BD}^F-\gamma)+(r_B r_D)^2\right)  +\nonumber \\ &&
\Akpi^2|\bwoth(s_{13})|^2 \left(r_D^2+2 r_B r_D
  \cos(\delta_{BF}^D-\gamma)+r_B^2\right)  + \nonumber \\ 
&& 2\Arhok \Akpi| \bwtt(s_{23})\bwoth(s_{13})| \times \nonumber\\ &&
\quad \Big\{  r_D \cos \delta_0^{D+} 
-  r_B^2 \cos \delta_0^{F+} 
-r_B  r_D \cos (\delta_{B0}^{D+}-\gamma)+ 
r_B \cos (\delta_0^{BF+}+\gamma)\Big\} + 
\nonumber\\
&& 2\Arhok \Akpi|\bwtt(s_{23})\bwotw(s_{12})| \times \nonumber\\ &&
\quad \Big\{\cos \delta_0^{F-}
- r_B \cos (\delta_{B0}^{F-} -\gamma) 
+r_B  r_D \cos (\delta_0^{BD-} +\gamma)
- r_B^2 r_D \cos \delta_0^{D-}\Big\} + \nonumber\\
&&2\Akpi^2 |\bwotw(s_{12})\bwoth(s_{13})| \times \nonumber\\ &&
\quad \Big\{ r_D \cos \delta_{F-}^{D+}
+  r_B \cos (\delta_-^{B+} +\gamma) 
+r_B  r_D^2 \cos (\delta_{B-}^+ -\gamma)
+ r_B^2 r_D \cos \delta_{D-}^{F+}\Big\},
\eeqa
where the notation of the strong phases is such that the lower (upper)
indices indicate phases appearing with a plus (minus) sign.
For example,
\beqa
\delta_{D-}^{F+} = \delta_D +\delta_- - \delta_F -
\delta_+.
\eeqa
$\Arhok$, $\Akpi$ and $r_D$ are assumed to be known and
thus there are  five unknowns to fit, namely
\beqa
r_B,\quad \delta_D,\quad \delta_F, \quad \delta_B,\quad \gamma\,.
\eeqa
Using both $B^-$ and $B^+$ decays, there is enough
information to determine them all.
This is true even if one neglects terms that scale as $r_B^2$ and
even if $r_D=0$. This indicates that the method does not rely on
doubly Cabibbo suppressed decays of the $D$, and that it is sensitive
to $\gamma$ in terms of order $r_B$, rather than $r_B^2$ (See
discussion in~\cite{Grossman:2002aq}).  Moreover, even if some or all of the strong phases that
arise from two-body decays, namely, $\delta_B$,
$\delta_D,$ and $\delta_F$, vanish, there is still enough information
to determine $\gamma$.

\section{Multibody $B$ decay}\label{app:multi}
We consider the cascade decay $B^-\to D X_s^- \to (K_S \pi^-\pi^+)_D X_s^-$.
Let us assume that the phase space of the first decay, $B^-\to D
X_s^-$, is partitioned into $m$ bins that we label by the index $j$, and
the phase space of the $D$ meson decay is partitioned into $n=2k$ bins
labeled by $i$ and $\bar{i}$ as in section \ref{model-indep}. Instead of
Eqs. \eqref{relations4k} we now have the set of $4k\times m$ equations
\begin{subequations}\label{relations4kxm}
\begin{align}
\begin{split}
\hat\Gamma^-_{i,j} \equiv 
\int_{i,j} d\Gamma(B^- \to (&K_S \pi^- \pi^+)_D X_s^-)=\\
&T_i +R_j^B T_{\overline{i}} + \cos\gamma (c_i c_j^B+s_is_j^B) +
\sin\gamma (c_is_j^B -s_i c_j^B),
\end{split}
\\
\begin{split}
\hat\Gamma^-_{\bar{i},j} \equiv 
\int_{\bar{i},j} d\Gamma(B^- \to (&K_S \pi^- \pi^+)_D X_s^-)=\\
&T_{\overline{i}} +R_j^B T_i + \cos\gamma (c_i c_j^B-s_is_j^B) +
\sin\gamma (c_is_j^B +s_i c_j^B),
\end{split}
\\
\begin{split}
\hat\Gamma^+_{i,j} \equiv 
\int_{i,j} d\Gamma(B^+ \to (&K_S  \pi^- \pi^+)_D X_s^+)=\\
&T_{\overline{i}} +R_j^B T_i + \cos\gamma (c_i c_j^B-s_is_j^B) -
\sin\gamma (c_is_j^B +s_i c_j^B),
\end{split}
\\
\begin{split}
\hat\Gamma^+_{\bar{i},j} \equiv 
\int_{\bar{i},j} d\Gamma(B^+ \to (&K_S  \pi^- \pi^+)_D X_s^+)=\\
&T_i +R_j^B T_{\overline{i}} + \cos\gamma (c_i c_j^B+s_is_j^B) -
\sin\gamma (c_is_j^B -s_i c_j^B),
\end{split}
\end{align}
\end{subequations}
where the integration is over the phase space of the $j$-th bin in the $B$
decay and the phase space of the $i$-th bin in the $D$ decay. The $j$-th bin
of the $B^+$ decay phase space is obtained from the $j$-th bin of the 
$B^-$ decay by $CP$ conjugation. 
We also used
\beqa
s_j^B &=& \int_j 2r_B\sin\delta_B
 ,\nonumber \\
c_j^B &=& \int_j
2r_B\cos\delta_B 
,\nonumber \\
R_j^B &=& \int_j r_B^2, 
\eeqa
where $r_B$ and $\delta_B$ are functions of the
position in the $B$ decay phase space. {}From the set of $4k\times m$
equations \eqref{relations4kxm}, one has to determine $2k+3m+1$
unknowns $c_i$, $s_i$, $c_j^B$, $s_j^B$, $R_j^B$, and $\gamma$. With a
partition of the $D$ decay phase space into $2 k\ge 4$ bins and with a
partition of the $B$ decay phase space into $m \ge 1$ bins, one has
enough relations to determine all the unknowns, including the angle
$\gamma$. This is true even  for constant $\delta_B$ and $r_B$, in
which case the
above equations fall into $4k$ sets of $m$ equivalent relations,
i.e. the set of $4k\times m$ equations is reduced to the set of $4k$
independent relations
\eqref{relations4k}.

Finally, we note that  the above equations can
be used to determine $\gamma$ also for two-body $D$ decays \cite{Aleksan:2002mh}.



\end{document}